\documentclass[12pt]{article}
\usepackage{epsfig}
\usepackage{color}

\usepackage{braket}

\newcommand{\be}{\begin{equation}}
\newcommand{\ee}{\end{equation}}

\newcommand{\bea}{\begin{eqnarray}}
\newcommand{\eea}{\end{eqnarray}}
\newcommand{\beq}{\begin{equation}}
\newcommand{\eeq}{\end{equation}}
\newcommand{\nn}{\nonumber}

\def\fun#1#2{\lower3.6pt\vbox{\baselineskip0pt\lineskip.9pt
\ialign{$\mathsurround=0pt#1\hfil##\hfil$\crcr#2\crcr\sim\crcr}}}

\begin{document}

\title{
Pentaquarks
and resonances in the $pJ/\psi$ spectrum
}
\author{
V.V. Anisovich$^+$, M.A. Matveev$^+$, J. Nyiri$^*$, A.V. Sarantsev$^{+ \diamondsuit}$,
\\
 A.N. Semenova$^+$
}

\date{}
\maketitle

\begin{center}
{\it
$^+$National Research Centre ''Kurchatov Institute'':
Petersburg Nuclear Physics Institute, Gatchina, 188300, Russia}

{\it $^\diamondsuit$
Helmholtz-Institut f\"ur Strahlen- und Kernphysik,
Universit\"at Bonn, Germany}

{\it $^*$Institute for Particle and Nuclear Physics, Wigner RCP,
Budapest 1121, Hungary}

\end{center}

\begin{abstract}
We consider exotic baryons with hidden charm as
antiquark-diquark-diquark composite systems. Spin and isospin structure
of such exotic states is given and masses are estimated.
The data for production of pentaquarks in the reaction $\Lambda_b\to K^-p J/\psi$ are discussed. We suggest that the narrow peak in $pJ/\psi$ spectra at 4450 MeV
is antiquark-diquark-diquark state with negative parity, $5/2^-(4450)$, while the broad bump
$3/2^+(4380)$ is the result of
rescatterings in the ($pJ/\psi$)-channel.
Positions of other pentaquarks with negative parity are estimated.

\end{abstract}

PACS:
12.40.Yx, 12.39.-x, 14.40.Lb

\section{Introduction}

Data for the decay $\Lambda^0_b\to p J/\psi K^-$ \cite{LHCbB}
provide a definite argument for the existence
of a pentaquark, a baryon system in $ p J/\psi$ spectra which has the following quark content:
 \be \label{1}
P^+_{\bar c cuud}=\bar c(cuud).
 \ee
In terms of the quark-diquark states it can be presented as a
three-body systems:
\be \label{2}
\bar c(cuud)= \bar c\cdot(cu)\cdot(ud) +{\rm permutations\, of\, the}\, u,d\,
{\rm quarks }.
\ee
A diquark is a color triplet member, similar to a quark, and the right-hand side of eq. (2)
presents a three-body system with a color structure similar to that in low-lying baryons. It
is reasonable to suppose that we face similar color forces in three quark and
antiquark-diquark-diquark systems as well. Following this idea we perform a
classification of such baryon states and give estimations of their masses.
Estimation of diquark masses is given in \cite{cq-cq} where diquark-antidiquark
states are studied.

The notion of the diquark was introduced
by Gell-Mann \cite{gell-mann}. Diquarks were discussed for baryon states
during a long time, see pioneering papers \cite{ida,licht,ono,vva75,schm} and conference
presentations \cite{diquark1,goeke,diquark2}. The systematization of baryons in terms of
the quark-diquark states is presented in \cite{qD,book4}.
The application of the diquarks
to exotic mesons in the sector of
heavy  diquarks was discussed by Maiani $et\,al.$ \cite{maiani},
Voloshin \cite{voloshin}, Ali $et\, al.$ \cite{ali}.

Pentaquarks built of light-light and light-heavy diquarks present natural extension
of multiquark schemes studed in the last decade
for mesons \cite{wein,brod} and baryons \cite{jaff,ross}.

\section{Pentaquarks}
In color space we write for the pentaquark:
\bea
&&
P^+_{\bar ccuud }=
\epsilon_{\alpha\beta\gamma}\bar c^\alpha(cu)^\beta(ud)^\gamma\,+\,
{\rm permutations\, of\, the}\, u,d\,{\rm quarks }
\quad,
\\
&&
(cu)^\beta=\epsilon^{\beta\beta'\gamma'}c_{\beta'}u_{\gamma'},
\quad
(ud)^\gamma=\epsilon^{\gamma\beta''\gamma''}u_{\beta''}d_{\gamma''},
\nn
\eea
where $\alpha,\beta,\gamma$ refer to color indices.

We discuss a scheme in which the exotic baryon states are formed by standard
QCD-motivated interactions (gluonic exchanges, confinement forces) but
in addition with diquarks as constituents.

\subsection{Spin structure of the pentaquarks}

We work with two diquarks: scalar $S$ and axial-vector $A$. In terms of these diquarks
the color-flavor wave funcion of pentaquark reads:
\be \label{4}
P_{\bar c\cdot (cq)\cdot(q'q'')}=\bar
c^\alpha\cdot \epsilon_{\alpha\beta\gamma}\,
\begin{tabular}{|l|}
$S^\beta_{(cq)}$\\
$A^\beta_{(cq)}$
\end{tabular}
\cdot
\begin{tabular}{|l|}
$S^\gamma_{(q'q'')}$\\
$A^\gamma_{(q'q'')}$
\end{tabular}
\ee
We have six diquark-diquark states:
\be          \label{5}
P_{\bar c\cdot (cq)\cdot(q'q'')}= \bar c^\alpha\cdot
\begin{tabular}{|l|}
$(S_{(cq)}S_{(q'q'')})^\alpha(0^{+})$\\
$(S_{(cq)}A_{(q'q'')})^\alpha(1^{+})$\\
$(A_{(cq)}S_{(q'q'')})^\alpha(1^{+})$\\
$(A_{(cq)}A_{(q'q'')})^\alpha(0^{+})$\\
$(A_{(cq)}A_{(q'q'')})^\alpha(1^{+})$\\
$(A_{(cq)}A_{(q'q'')})^\alpha(2^{+})$
\end{tabular}
\ee
with the spin-parity numbers $J^P= 0^+, 1^+,2^+$.

\subsection{Isospin structure of the diquarks}

We face the following isospin states for the
diquarks:
\bea         \label{6}
&&
S_{(cq)}(I_d=1/2,\,J_d=0),\qquad A_{(cq)}(I_d=1/2,\,J_d=1),
\\
&&
S_{(q'q'')}(I_d=0,\,J_d=0),\qquad\,\, A_{(q'q'')}(I_d=1,\,J_d=1).
\nn
\eea
So, the isospin-spin sector of the pentaquarks $P(I,J^P)$ reads:
\be\label{7}
P_{\bar c\cdot (cq)\cdot(q'q'')}= \bar c^\alpha\cdot
\begin{tabular}{|l|}
$(S_{(cq)}S_{(q'q'')})^\alpha(0^{+})$\\
$(S_{(cq)}A_{(q'q'')})^\alpha(1^{+})$\\
$(A_{(cq)}S_{(q'q'')})^\alpha(1^{+})$\\
$(A_{(cq)}A_{(q'q'')})^\alpha(0^{+})$\\
$(A_{(cq)}A_{(q'q'')})^\alpha(1^{+})$\\
$(A_{(cq)}A_{(q'q'')})^\alpha(2^{+})$
\end{tabular}
           \Longrightarrow
\begin{tabular}{l}
$P(\frac12,\frac12^{-})$\\
$P(\frac12,\frac12^{-})$, $P(\frac12,\frac32^{-})$,
$P(\frac32,\frac12^{-})$, $P(\frac32,\frac32^{-})$, \\
$P(\frac12,\frac12^{-})$, $P(\frac12,\frac32^{-})$\\
$P(\frac12,\frac12^{-})$, $P(\frac32,\frac 12^{-})$\\
$P(\frac12,\frac12^{-})$, $P(\frac12,\frac32^{-})$, $P(\frac32,\frac12^{-})$,
$P(\frac32,\frac32^{-})$
\\
$P(\frac12,\frac32^{-})$, $P(\frac12,\frac52^{-})$,
$P(\frac32,\frac32^{-})$, $P(\frac32,\frac52^{-})$
\end{tabular}
 \ee

\subsection{Pentaquarks, their masses and spins}

Mass formula for a diquark-diquark system $(cq)\cdot (q'q'')$,
 is accepted to be the same as for the
diquark-antidiquark one \cite{cq-cq}. We write
\be  \label{8}
M_{(cq)\cdot (q'q'')}=m_{(cq)}+m_{(q'q'')}+
J_{(cq)\cdot (q'q'')}\,(J_{(cq)\cdot (q'q'')}+1)\,\Delta
\ee
with the parameters which were determined in ref. \cite{cq-cq}:
\bea       \label{9}
&&
\Delta=70\pm 10\mbox{ MeV},
\\ \nn
&&
m_{S_{(cq)}}=2000\pm 50 \mbox{ MeV} ,\qquad m_{A_{(cq)}}=2050\pm 50 \mbox{ MeV}.
\nn
\eea
Here we concentrate our attention on non-strange diquarks only,
for an expansion of the results to the strange quark sector we
need $m_{S_{(cs)}}$ and $m_{A_{(cs)}}$.

For the $(q'q'')$ diquarks we accept masses found in the analysis of
baryons \cite{qD,book4}:
\be\label{10}
m_{S_{(q'q'')}}=650\pm 50 \mbox{ MeV} ,\qquad m_{A_{(q'q'')}}=750\pm 50 \mbox{ MeV} .
\ee
The mass of the costituent antiquark $\bar c$ is equal to \cite{ADMNS-cc,mano}:
\be         \label{11}
m_c=1300\pm 50\mbox{ MeV}.
\ee

Within these masses (\ref{9}),(\ref{10}),(\ref{11}) we roughly estimate the masses
of the  pentaquarks as:
\be                  \label{12}
M_{\bar c\cdot (cq)\cdot (q'q'')}\simeq m_{\bar c}+M_{(cq)\cdot (q'q'')}\,.
\ee
Correspondingly we write a set of the low-laying pentaquark states:
\be                           \label{13}
P_{\bar c\cdot (cq)\cdot(q'q'')}=
\begin{tabular}{l|l}
           $\qquad I=1/2$&       $\qquad I=3/2 $                          \\
\hline
$P_{\bar c\,S_{(cq)}S_{(q'q'')}}^{(\frac12,\frac12^{-})}(3800)$, & \\
$P_{\bar c\,S_{(cq)}A_{(q'q'')}}^{(\frac12,\frac12^{-})}(4190)$,
$P_{\bar c\,S_{(cq)}A_{(q'q'')}}^{(\frac12,\frac32^{-})}(4190)$, &
$P_{\bar c\,S_{(cq)}A_{(q'q'')}}^{(\frac32,\frac12^{-})}(4190)$,
$P_{\bar c\,S_{(cq)}A_{(q'q'')}}^{(\frac32,\frac32^{-})}(4190)$, \\
$P_{\bar c\,A_{(cq)}S_{(q'q'')}}^{(\frac12,\frac12^{-})}(4140)$,
$P_{\bar c\,A_{(cq)}S_{(q'q'')}}^{(\frac12,\frac32^{-})}(4140)$, &\\
$P_{\bar c\,A_{(cq)}A_{(q'q'')}}^{(\frac12,\frac12^{-})}(4100)$, &
$P_{\bar c\,A_{(cq)}A_{(q'q'')}}^{(\frac32,\frac12^{-})}(4100)$,  \\
$P_{\bar c\,A_{(cq)}A_{(q'q'')}}^{(\frac12,\frac12^{-})}(4240)$,
$P_{\bar c\,A_{(cq)}A_{(q'q'')}}^{(\frac12,\frac32^{-})}(4240)$, &
$P_{\bar c\,A_{(cq)}A_{(q'q'')}}^{(\frac32,\frac12^{-})}(4240)$,
$P_{\bar c\,A_{(cq)}A_{(q'q'')}}^{(\frac32,\frac32^{-})}(4240)$, \\
$P_{\bar c\,A_{(cq)}A_{(q'q'')}}^{(\frac12,\frac32^{-})}(4520)$,
$P_{\bar c\,A_{(cq)}A_{(q'q'')}}^{(\frac12,\frac52^{-})}(4520)$, &
$P_{\bar c\,A_{(cq)}A_{(q'q'')}}^{(\frac32,\frac32^{-})}(4520)$,
$P_{\bar c\,A_{(cq)}A_{(q'q'')}}^{(\frac32,\frac52^{-})}(4520)$\,.
\end{tabular}
 \ee
Masses are given in MeV units, an uncertainty in the determination of masses
is of the order of $\pm$150 MeV.

\section{Discussion}

The state $P_{\bar c\,A_{(cq)}A_{(q'q'')}}^{(\frac12,\frac52^{-})}(4520\pm 150)$ is a good candidate
to be a state which was observed in \cite{LHCbB}: $5/2^?(4450\pm 4)$ with a width of
$\Gamma=39\pm 24$ MeV. Then the broad state,  $3/2^?(3380\pm 38)$
with $\Gamma=205\pm 94$ MeV also observed
in the $pJ/\psi$ spectrum, is a positive parity state, $3/2^+$.

An opposite classification of states is suggested in
\cite{mpr}: $3/2^?(3380\pm 38)\to 3/2^-$($S$-wave pentaquark) and
$5/2^?(4450\pm 4)\to 5/2^+$($P$-wave pentaquark).

We suppose that the broad bump in the $3/2^+$-wave is the result of rescatterings
in $pJ/\psi$-channel, for example, such as $\Lambda_b \to \Lambda(1520)J/\psi \to K^-
(pJ/\psi) $.

In the suggested scheme the mass interval $(4040-4500)$ MeV should contain additionally
several resonances with  $J^P=1/2^-,3/2^-$. Some of them may dominantly decay into
$p\eta_c$-channel,
concerning mainly the $1/2^-$ states.

The masses of pentaquarks with strange diquarks $(cs)$, see eq.(\ref{9}), are estimated
quite similar to the non-strange ones.

In conclusion, the scheme of the low-lying pentaquark is suggested
on the basis of the
study of the tetraquark states \cite{cq-cq}.
We are convinced that the crossing studies of exotic mesons and
baryons give a correct way for investigation of these topics.

The work was supported by grants RSGSS-4801.2012.2., RFBR-13-02-00425 and RSCF-14-22-00281 .

\end{document}